\documentclass[aps,preprint,showpacs]{revtex4}
\usepackage{amsfonts}
\usepackage{amsmath}
\usepackage{amssymb}
\usepackage{graphicx}

\begin{document}

\title{Dynamics of matter-wave solitons in a time-modulated two-dimensional
optical lattice.}
\author{Gennadiy Burlak$^{1}$ and Boris A. Malomed$^{2}$}
\affiliation{$^{1}$Centro de Investigaci\'{o}n en Ingenier\'{\i}a y Ciencias Aplicadas,
Universidad Aut\'{o}noma del Estado de Morelos, Cuernavaca, Mor., Mexico}
\affiliation{$^{2}$Department of Physical Electronics, School of Electric Engineering,
Faculty of Engineering, Tel Aviv University, Tel Aviv 69978, Israel}

\begin{abstract}
By means of the variational approximation (VA) and systematic simulations,
we study dynamics and stability boundaries for solitons in a two-dimensional
(2D) self-attracting Bose-Einstein condensate (BEC), trapped in an optical
lattice (OL) whose amplitude is subjected to the periodic time modulation
(the modulation frequency, $\omega $, may be in the range of several KHz).
Regions of stability of the solitons against the collapse and decay are
identified in the space of the model's parameters. A noteworthy result is
that the stability limit may reach the largest ($100\%$) modulation depth,
and the collapse threshold may exceed its classical value in the static
lattice (which corresponds to the norm of Townes soliton). Minimum norm $%
N_{\min }$ necessary for the stability of the solitons is identified too. It
features a strong dependence on $\omega $\ at a low frequencies, due to a
resonant decay of the soliton. Predictions of the VA are reasonably close to
results of the simulations. In particular, the VA helps to understand
salient resonant features in the shape of the stability boundaries observed
with the variation of $\omega $.
\end{abstract}

\maketitle


\section{Introduction}

A challenging subject in the study of dynamical patterns in Bose-Einstein
condensates (BECs) is the investigation of matter-wave solitons in
multidimensional settings. Various routes leading to the creation of stable
2D and 3D solitons have been elaborated theoretically. The proposed
approaches have much in common with their counterparts developed for the
stabilization of 2D and 3D spatiotemporal solitons in nonlinear optics, see
review \cite{Review}. However, thus far, only effectively one-dimensional
(1D) matter-wave solitons have been created in experiments which used
cigar-shaped traps to confine the condensate \cite{experiment} (the actual
shape of the soliton may be nearly three-dimensional, but it is
intrinsically self-trapped only in the longitudinal direction). A phenomenon
which impedes the stabilization of truly multidimensional solitons is the
occurrence of collapse in 2D and 3D condensates with attraction between
atoms. As demonstrated in Refs. \cite{BBB,BBB2,BBB3}, a universal method for
the stabilization of matter-wave solitons in the multidimensional setting
may be provided by optical lattices (OLs), i.e., periodic potentials which
are induced, through interference patterns, by coherent laser beams
illuminating the condensate in opposite directions (in principle, similar
methods may be applied to a gas of polaritons, where an evidence of the BEC
state was recently reported too \cite{Kasprzak:2006a}). The stabilization of
2D and 3D solitons is possible with the help of the fully-dimensional OL,
whose dimension is equal to that of the entire space, and by low-dimensional
lattices, whose dimension is smaller by one. The latter means quasi-1D and
quasi-2D OLs in the 2D and 3D space, respectively \cite{BBB2,Barcelona}. OLs
may also support solitons of a completely different type, namely, \textit{%
gap solitons} in BEC (of any dimension) with repulsion between atoms. In
this case, the solitons emerge due to the interplay between the repulsive
nonlinearity and the negative effective mass in parts of the linear bandgap
spectrum generated by the OL \cite{gap1D}. Although gap solitons cannot
represent the ground state of the respective system, they are dynamically
stable objects \cite{GSstability}. The creation of effectively 1D gap
solitons, composed of $\simeq 250$ atoms of $^{87}$Rb, was reported in Ref.
\cite{Eiermann:2004a}, see also review \cite{OliverMorsch:2006a}.
Multidimensional (chiefly, 2D) gap solitons \cite{gap2D}, and ``semi-gap"
solitons (which look like gap solitons in one direction, and regular
solitons in the other \cite{semi-gap}) were predicted too \cite{gap2D}, but
not yet observed in the experiment. Stable gap solitons, in 1D and 2D
settings alike, can be supported not only by periodic OLs, but also by
quasi-periodic ones \cite{HidetsuguSakaguchi:2006a}.

It is relevant to note that the presence of the lattice, defining a
particular spatial scale in the system (the OL period), introduces
an effective nonlocality. On the other hand, a nonlocality can be
directly induced by long-range interactions between atoms carrying
dipolar moments. In this way, it has been demonstrated that the
interplay between the local repulsion and long-range attraction may
also support
stable 2D solitons, both isotropic \cite{DDiso} and anisotropic \cite%
{DDaniso} ones.

Another theoretically elaborated stabilization technique relies on periodic
time modulation of the effective nonlinearity coefficient, which may be
provided by the Feshbach-resonance effect in a low-frequency ac magnetic
field applied to the condensate. It has been demonstrated that this method
can stabilize 2D fundamental solitons, but not 3D ones \cite{FRM} (the
stabilization of 3D solitons and their bound complexes is possible under a
combined action of this technique and a quasi-1D OL \cite{Warsaw}; the
application of the Feshbach-resonance technique to matter waves trapped in
usual OLs \cite{Mason}, as well as in parabolic traps \cite{Fatkhulla}, was
studied too). This approach to the creation of stable 2D matter-wave
solitons followed the earlier elaborated mechanism providing for the
stabilization of (2+1)-dimensional optical spatial solitons in the model of
a bulk medium built as a periodic concatenation of layers with self-focusing
and self-defocusing Kerr nonlinearity \cite{Isaac}. It should be noted,
however, that very accurate simulations of the so stabilized 2D solitons in
the framework of the respective Gross-Pitaevskii equation (GPE) demonstrate
that they may be subject to an extremely slow decay \cite{Itin}.

The methods using the periodic modulation of the nonlinearity belong to a
broad class of techniques relying on the \textit{periodic management} of
solitons \cite{book}. In terms of the BEC in one dimension, another relevant
example representing this class is the \textit{lattice management} in the
model where the strength of the OL is subjected to periodic time modulation
\cite{Dalfovo,Thawatchai} (in the experiment, this was used as a method to
excite the condensate \cite{experiment-OL}). The main result of the analysis
of the model including the lattice management in combination with the
self-repulsive nonlinearity, reported in Ref. \cite{Thawatchai}, was
identification of stability limits for several families of gap solitons in
this setting (fundamental solitons, double-hump bound states, etc.). Also
belonging to this general class are settings based on the periodic
modulation in time of the strength of a parabolic potential trap which
confines the condensate \cite{trap-modulation}, and periodic modulation of
the transverse trapping frequency which supports cigar-shaped BEC
configurations. In the latter case, the ``management" may induce Faraday
waves in the condensate, as was predicted theoretically \cite{Faraday-theory}
and demonstrated experimentally \cite{Engels}. It is also relevant to
mention that another variety of the "management" was proposed very recently,
with the aim to stabilize 2D matter-wave solitons: driven Rabi oscillations
in a mixture of two mutually interconvertible BEC species (different states
of the same atom), with self-attractive and self-repulsive intra-species
interactions \cite{Randy}.

\section{The model and outline of the paper}

A natural extension of the studies outlined above is the application of the
lattice-management technique to multidimensional solitons. This topic
includes several distinct problems, depending on the sign of the intrinsic
nonlinearity. In the 1D setting, the lattice management for solitons in the
attraction model is not a fundamentally important issue, because stable 1D
solitons in the self-attractive condensate exist without any lattice \cite%
{experiment}. However, the model with the intrinsic attraction poses a new
problem in the 2D case, as, in the absence of the lattice, 2D solitons are
unstable against the collapse, and, as said above, the simplest way to
suppress the instability is provided by the use of the quasi-1D OL \cite%
{BBB2}. Therefore, it may be interesting to identify stability limits for 2D
solitons in the attractive model with the quasi-1D lattice whose strength is
subjected to the periodic time modulation; in particular, one may be
wondering if the solitons may survive in the case when the modulation depth
attains its maximum, so that the lattice periodically vanishes. This problem
was a subject of recent work \cite{Thawatchai2}, which was also dealing with
collisions between the 2D solitons, that may move freely in the unconfined
direction. It was found that the actual stability region is quite limited:
typically, the solitons loose their stability before the modulation depth
attains $50\%$.

The objective of the present work is to construct soliton families and
identify their stability boundaries in the 2D attractive model with the full
2D lattice subject to the time-periodic modulation. In the normalized form,
the respective two-dimensional GPE for mean-field wave function $\Psi \left(
x,y,t\right) $ is%
\begin{equation}
i\frac{\partial \Psi }{\partial t}=-\frac{1}{2}\left( \frac{\partial ^{2}}{%
\partial x^{2}}+\frac{\partial ^{2}}{\partial y^{2}}\right) \Psi -\left\vert
\Psi \right\vert ^{2}\Psi -V_{0}\left[ 1+\frac{\varepsilon }{2}\cos (\omega
t)\right] \left[ \cos \left( 2x\right) +\cos \left( 2y\right) \right] \Psi ,
\label{GPE}
\end{equation}%
where $t$ is time, $\left( x,y\right) $ are coordinates in the 2D plane
(scaled so as to fix the OL period equal to $\pi $), and $V_{0}$ is the
strength of the lattice, while $\varepsilon $ and $\omega $ are the
amplitude and frequency of its temporal modulation. Coefficient $-1$ in
front of the nonlinear term in Eq. (\ref{GPE}) implies that the nonlinearity
is attractive. In the case of the quasi-1D lattice \cite{BBB2,Thawatchai2},
term $\cos (2y)$ is dropped from the potential in Eq. (\ref{GPE}). Actually,
$V_{0}$ is measured in units of the recoil energy of atoms trapped in the
OL. For a typical case of atoms of $^{7}$Li loaded in the OL with the period
on the order of $\mu $m, a characteristic value of the scaled frequency, $%
\omega \sim 2$ (see below), corresponds, in physical units, to the
modulation rate on the order of several KHz.

An experimental implementation of the model may be provided by periodic
attenuation of the intensity of the laser beams that create the OL. Since
this management mode cannot change the sign of the OL's amplitude, the
modulation coefficient in Eq. (\ref{GPE}) is restricted to interval $0\leq
\varepsilon \leq 2$. We, chiefly, follow this restriction below. On the
other hand, Eq. (\ref{GPE}) with the modulated potential that formally
corresponds to $\varepsilon \rightarrow \infty $, \textit{viz}., $-V_{0}\cos
(\omega t)\left[ \cos \left( 2x\right) +\cos \left( 2y\right) \right] $, may
be realized as a superposition of several moving OLs \cite{Staliunas:2007a}
(a 1D counterpart of that setting was introduced in Ref. \cite%
{Staliunas:2006a}). In works \cite{Staliunas:2006a} and \cite%
{Staliunas:2007a} it was demonstrated that such potentials, if combined with
the self-repulsive nonlinearity, give rise to anomalously narrow \textit{%
subdiffractive} gap solitons, for which the usual second-order operator of
the kinetic energy is, effectively, replaced by a fourth-order operator. On
the other hand, a piecewise-constant time-periodic modulation mode was
considered, for both periodic and quasi-periodic 2D potentials, in recent
work \cite{Gena} (also in the combination with the self-repulsive
nonlinearity). The result was generation of a multi-soliton lattice,
starting from an initial localized state.

The rest of the paper is organized as follows. In Section III, we
elaborate the variational approximation (VA) for the analysis of the
dynamics of fundamental 2D solitons in the model based on Eq.
(\ref{GPE}). In Section IV, results of systematic direct simulations
of the soliton evolution in Eq. (\ref{GPE}) are reported, and
compared with predictions of the VA. The results are summarized in
the form of stability regions for the solitons in the model with the
time-modulated OL. The most noteworthy finding reported in Section
IV is that, depending on $\omega $, the solitons may remain stable
up to the maximum ($100\%$) modulation depth, in contrast to the
shallow modulation that limited the stability of the solitons
supported by the quasi-1D OL \cite{Thawatchai2}. It is also worthy
to mention a multi-resonance shape featured by the stability
boundary in the $\left( \omega ,\varepsilon \right) $ plane, and an
increase of the collapse threshold (defined in terms of the
soliton's norm) provided by the time-modulated OL, in comparison
with the static lattice, i.e., the threshold may exceed the norm of
the Townes soliton. The minimum norm necessary for the stability of
the solitons, $N_{\min }$, is identified too, as a function of
$\omega $ and $\varepsilon $ (this characteristic was considered
earlier only in the static model \cite{Salerno}). Due to the
possibility of a resonantly enhanced decay of the soliton, $N_{\min
}$ features a strong dependence on $\omega $ at low frequencies. The
paper is concluded by Section V.

\section{The variational approximation}

\subsection{ Effective Lagrangian}

Variational methods have been quite useful in many problems of nonlinear
optics and BEC \cite{Desaix,VA,BBB,BBB2,FRM,book}. To apply the VA to the
present model, we notice that Eq. (\ref{GPE}) can be derived from Lagrangian
$L=\int_{-\infty }^{+\infty }\mathcal{L}dx$, with density%
\begin{align}
\mathcal{L}& =\frac{i}{2}\left( \Psi ^{\ast }\Psi _{t}-\Psi \Psi _{t}^{\ast
}\right) -\frac{1}{2}\left( \left\vert \Psi _{x}\right\vert ^{2}+\left\vert
\Psi _{y}\right\vert ^{2}\right) +\frac{1}{2}\left\vert \Psi \right\vert ^{4}
\notag \\
& +V_{0}\left[ 1+\frac{\varepsilon }{2}\cos (\omega t)\right] \left[ \cos
\left( 2x\right) +\cos \left( 2y\right) \right] \left\vert \Psi \right\vert
^{2},  \label{L}
\end{align}%
where the asterisk stands for the complex conjugation. Following Refs. \cite%
{BBB} and \cite{BBB2}, we adopt the isotropic \textit{ansatz} for the
soliton,%
\begin{equation}
\Psi _{\mathrm{ans}}\left( x,y,t\right) =A(t)\exp \left( i\phi (t)+\frac{i}{2%
}b(t)r^{2}-\frac{r^{2}}{2W^{2}(t)}\right) ,  \label{ans}
\end{equation}%
where $r^{2}\equiv x^{2}+y^{2}$, and all variables $A(t),$ $\phi (t),$ $%
b(t), $ and $W(t)$ (amplitude, phase, radial \textit{chirp}, and radial
width, respectively) are real. Tractable, although rather cumbersome,
variational equations may also be generated by an anisotropic generalization
of ansatz (\ref{ans}), with different widths and chirps along axis $x$ and $%
y $. However, subsequent analysis of those equations does not predict stable
anisotropic solitons.

The substitution of the ansatz in Eq. (\ref{L}) and calculation of the
integrals yield the effective Lagrangian,%
\begin{align}
L_{\mathrm{eff}}& =-N\frac{d\phi }{dt}-\frac{N}{2W^{2}}+\frac{N^{2}}{4\pi
W^{2}}+2V_{0}\left[ 1+\frac{\varepsilon }{2}\cos (\omega t)\right]
Ne^{-W^{2}}  \notag \\
& -\frac{1}{2}\frac{db}{dt}NW^{2}-\frac{1}{2}b^{2}NW^{2},  \label{Leff}
\end{align}%
where $N\equiv \pi A^{2}W^{2}$. The first Euler-Lagrange equation following
from effective Lagrangian (\ref{Leff}), $\delta \left( \int L_{\mathrm{eff}%
}dt\right) /\delta \phi =0$ ($\delta /\delta \phi $ stands for the
variational derivative of the action functional), is tantamount to the
conservation of the norm of the wave function, which is the single dynamical
invariant of Eq. (\ref{GPE}). Indeed, the norm of ansatz (\ref{ans}) is
\begin{equation}
\int \int \left\vert \Psi _{\mathrm{ans}}(x,y)\right\vert ^{2}dxdy=\pi
A^{2}W^{2}\equiv N.  \label{N}
\end{equation}%
The second Euler-Lagrange equation, $\delta \left( \int L_{\mathrm{eff}%
}dt\right) /\delta b=0$, reduces to the well-known expression for the chirp
in terms of the time derivative of the width \cite{Desaix,VA}, $%
b=W^{-1}\left( dW/dt\right) $. Using this relation, the next variational
equation, which accounts to $\partial L_{\mathrm{eff}}/\partial \left(
W^{2}\right) =0$ [since Lagrangian (\ref{Leff}) does not contain $dW/dt$]
can be cast in the following final form,
\begin{equation}
\frac{d^{2}W}{dt^{2}}=\frac{1-N/\tilde{N}_{\max }}{W^{3}}-4V_{0}\left[ 1+%
\frac{\varepsilon }{2}\cos (\omega t)\right] W\exp \left( -W^{2}\right) ,
\label{W}
\end{equation}%
where $\tilde{N}_{\max }\equiv 2\pi $ is the well-known VA prediction \cite%
{Desaix} for the critical (maximum) norm in the 2D space, which separates
collapsing solutions at $N>\tilde{N}_{\max }$, i.e., ones with $%
W(t)\rightarrow 0,~A(t)\rightarrow \infty $ at $t\rightarrow t_{\mathrm{%
collapse}}$ [for $V_{0}=0$ and initial conditions $W(t=0)=W_{0}$ and $%
dW/dt(t=0)=0$, the collapse time predicted by Eq. (\ref{W}) is $t_{\mathrm{%
collapse}}=W_{0}^{2}\left( N/\tilde{N}_{\max }-1\right) ^{-1/2}$], and
noncollapsing ones at $N<\tilde{N}_{\max }$.

\subsection{Analysis of variational equation.}

The actual maximum value of $N$, found numerically from Eq. (\ref{GPE}) with
$V_{0}=0$ (it gives the norm of the \textit{Townes soliton} \cite{Berge}),
is
\begin{equation}
N_{\max }=5.85\approx \allowbreak 0.93\tilde{N}_{\max },  \label{Nmax}
\end{equation}%
which characterizes the accuracy of the VA. For the condensate of $^{7}$Li
atoms, a typical collapse threshold corresponds to the number of atoms $%
\lesssim 10^{4}$.

Equation (\ref{W}) helps one to understand what may happen to the 2D soliton
under the action of the weak \textquotedblleft management", with $%
\varepsilon /2\ll 1$. First, for $\varepsilon =0$, Eq. (\ref{W}) predicts a
stable equilibrium position, which is given by a smaller root of equation%
\begin{equation}
W_{0}^{4}\exp \left( -W_{0}^{2}\right) =\frac{1-N/\tilde{N}_{\max }}{4V_{0}}
\label{W0}
\end{equation}%
(the larger root gives an unstable solution). Then, the linearization of Eq.
(\ref{W}) (still with $\varepsilon =0$) yields the eigenfrequency of small
oscillations around $W=W_{0}$,%
\begin{equation}
\omega _{0}=\frac{\sqrt{2\left( 1-N/\tilde{N}_{\max }\right) \left(
2-W_{0}^{2}\right) }}{W_{0}^{2}}.  \label{omega0}
\end{equation}%
For instance, in the example considered in the next section (see Fig. \ref%
{Pic_A_max_w} below), with $1-N/\tilde{N}_{\max }=0.155$ and $V_{0}=0.25$,
the relevant root of Eq. (\ref{W0}) is $W_{0}\approx 0.71$, and Eq. (\ref%
{omega0}) yields $\omega _{0}\approx 1.35$. Keeping quadratic and cubic
terms in the expansion of Eq. (\ref{W}) in powers of $w(t)\equiv
W(t)/W_{0}-1 $ around $W$ leads to a standard equation of driven nonlinear
oscillations. In particular, in the near-critical situation, i.e., for $1-N/%
\tilde{N}_{\max }\ll 1$, this equation takes the form of
\begin{equation}
\frac{d^{2}w}{dt^{2}}+16V_{0}w-24V_{0}w^{2}+40V_{0}w^{3}=-2\varepsilon
V_{0}\cos (\omega t)-2\varepsilon V_{0}\cos \left( \omega t\right) w.
\label{simple}
\end{equation}%
It predicts the lowest-order direct resonance when $\omega $ is close to $%
\omega _{0},$ the parametric resonance at $\omega $ close to $2\omega _{0}$,
and higher-order resonances at $\omega =n\omega _{0}$, with $n=2,3,4,..$
\cite{Landau}. The resonances may help to \emph{stabilize} the 2D soliton
against the collapse, as, increasing the amplitude of its intrinsic
oscillations, the soliton spends less time in the \textquotedblleft
dangerous zone" with small width, that might be a starting point for the
collapse. On the other hand, effectively stretching the soliton, the
resonances may \emph{destabilize} it against decay into radiation, at
smaller values of $N$. Both trends are observed in numerical simulations, as
shown below. Comparison of predictions following from a numerical solution
of full variational equation (\ref{W}) and direct simulations of Eq. (\ref%
{GPE}) is presented in the next section.

\section{Numerical results}

\subsection{The simulation mode}

\begin{figure}[tbp]
\begin{center}
\includegraphics[height=4.3785in, width=6.2396in]{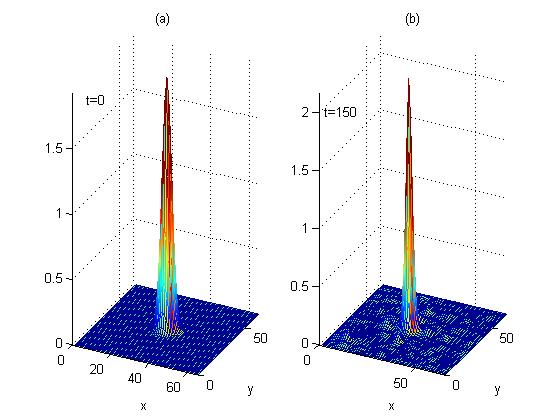}
\end{center}
\caption{(Color online) A typical example of a stable 2D soliton, as
obtained from the numerical solution of Eq.(\protect\ref{GPE}) with $%
V_{0}=0.65$, $\protect\varepsilon =0.5$, $\protect\omega =1.35$, and initial
configuration (\protect\ref{Init state}) with $A_{0}=1.39$, $q=0.5$, whose
norm, $N=6.07$, \emph{exceeds} the collapse threshold in the static model, $%
N_{\max }\approx 5.85$. The panels display the density distribution, $%
\left\vert \Psi (x,y,t)\right\vert ^{2}$, at $t=0$ (a) and $t=150$ (b).}
\label{Pic_OneSol_PL}
\end{figure}

Systematic simulations of Eq. (\ref{GPE}) were performed by means of the
split-step method \cite{Press}, in the $\left( x,y\right) $ domain of size $%
256\times 256$ or $512\times 512$. The simulations were run with the
Gaussian initial configuration,
\begin{equation}
\Psi (x,y,0)=A_{0}\exp \left( -q\left[ \left( x-x_{0}\right) ^{2}+\left(
y-y_{0}\right) ^{2}\right] \right) ,~q>0,  \label{Init state}
\end{equation}%
whose norm is $N=\pi A_{0}^{2}/q$ [cf. ansatz (\ref{ans})]. Taking $%
N<N_{\max }$, as well as $N$ slightly \emph{exceeding} $N_{\max }$ [recall $%
N_{\max }$, the critical value of the norm for the onset of the free-space
collapse, is given by Eq. (\ref{Nmax})], it was quite easy to find stable
solitons which keep their shape despite the temporal modulations imposed by
the ``management", see generic examples in Figs. \ref{Pic_OneSol_PL} and \ref%
{Pic_sinhron}. The former figure presents comparison of the initial and
final shapes of the soliton, while the latter one displays the evolution of
the soliton's amplitude in the course of its self-adjustment to the stable
shape.

\begin{figure}[tbp]
\begin{center}
\includegraphics[height=4.3785in, width=6.2396in]{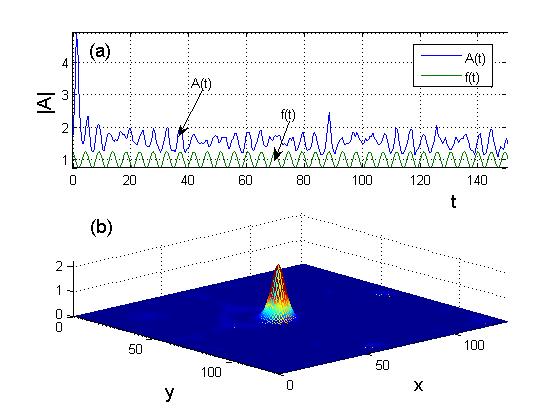}
\end{center}
\caption{(Color online) (a) The evolution of the amplitude of the stable
soliton, $|A|\equiv \left\vert \Psi \left( x=x_{0},y=y_{0}\right)
\right\vert $, for the same parameters as in Fig. \protect\ref{Pic_OneSol_PL}%
, except for $V_{0}=0.5$ ($|A|$ has the same meaning in other figures).
Curve $f(t)$ shows the modulation function in Eq.(\protect\ref{GPE}), $%
1+\left( \protect\varepsilon /2\right) \sin \left( \protect\omega t\right) $%
; (b) The shape of the soliton at $t=150$.}
\label{Pic_sinhron}
\end{figure}


The soliton shapes displayed in Figs. \ref{Pic_OneSol_PL} and \ref%
{Pic_sinhron} are confined, essentially, to a single cell of the OL (similar
to results reported in previous works \cite{BBB,BBB3,Barcelona}). Roughly
the same shapes would be observed in a parabolic trapping potential;
however, the difference is that, if the nonlinearity is too weak, the OL
cellular potential cannot suppress the tunnel decay of the localized pulses
\cite{Salerno}, therefore the decay is observed at smaller values of $N$, as
shown below.

\begin{figure}[tbp]
\begin{center}
\includegraphics[height=4.3785in, width=6.2396in]{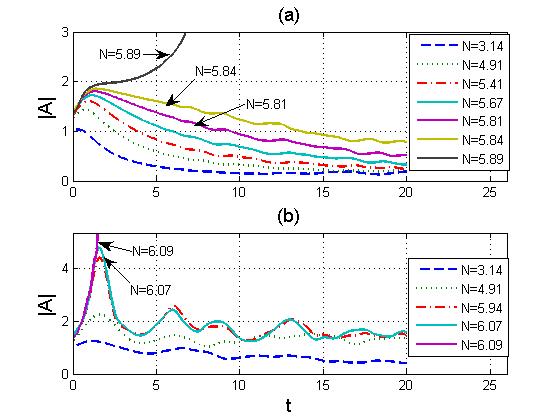}
\end{center}
\caption{(Color online) The evolution of the soliton's amplitude, as
obtained from the numerical solution of Eq. (\protect\ref{GPE}) for $\protect%
\varepsilon =0.5$ and $\protect\omega =1$ and different values of norm $N$
of initial configuration (\protect\ref{Init state}) (higher curves
correspond to larger $N$). (a) The model without the optical lattice, $%
V_{0}=0$. In this case, all configurations with $N<N_{\max }\approx 5.85$
decay, while the ones with $N>N_{\max }$ suffer the collapse. (b) In the
presence of the time-modulated optical lattice with $V_{0}=0.5$, the 2D
solitons may be stable, including some values of $N$ \emph{above} $N_{\max }$%
. In particular, the soliton is stable at $N=6.01\approx 1.03N_{\max }$,
while it collapses for $N=6.09$, at $t=1.49$. If $N$ is too small, $%
N<N_{\min }$, the soliton gradually decays into radiation, see Fig. \protect
\ref{Pic_n_MaxMin_V0} below. In panel (b), the curve for $N=3.14$, which is
slightly smaller than the corresponding value of $N_{\min }$, also shows the
slow decay of the soliton.}
\label{pic_A_V0_t}
\end{figure}

As said above, in the 2D equation with $V_{0}=0$ localized configurations
with $N>N_{\max }=5.85$ suffer collapse, while those with $N<N_{\max }$
decay. The static OL stabilizes 2D solitons in the latter case, but it
cannot arrest the collapse of initial localized states with $N>N_{\max }$
\cite{BBB,BBB2,BBB3}. The numerical analysis of the model with the quasi-1D
lattice potential subjected to the periodic time modulation did not reveal
stable solitons with $N>N_{\max }$ either \cite{Thawatchai2}. As shown in
Fig. \ref{pic_A_V0_t}, in the present model the periodic time modulation of
the two-dimensional OL potential makes it possible to stabilize the solitons
both at $N<N_{\max }$ and in some interval \emph{above} $N_{\max }$. The
actual increase of critical norm is not large, but the very fact that the
constraint $N\leq N_{\max }$ can be broken is an interesting result, as it
has never been reported before, to the best of our knowledge.


\subsection{Comparison with the variational approximation}

\begin{figure}[tbp]
\begin{center}
\includegraphics[height=4.3785in, width=6.2396in]{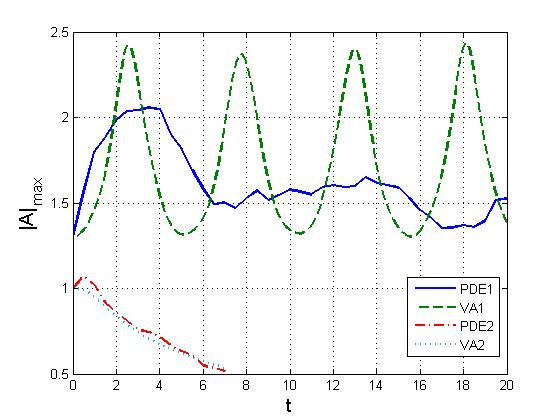}
\end{center}
\caption{(Color online) Comparison of the evolution \ of the soliton's
amplitude, as predicted by the variational approximation (``VA") and found
from direct simulations of the partial differential equation (\protect\ref%
{GPE}) (``PDE"), for $N=5.4$ (a stable soliton) and $N=4.6$ (a decaying
state).}
\label{Pic_va}
\end{figure}

Comparison of the predictions of the VA with numerical results in two
typical cases (for stable and decaying solitons, corresponding to $N=5.4$
and $N=4.6$, respectively) is presented in Fig. \ref{Pic_va}. The
discrepancy in the dependence of the soliton's amplitude on time, observed
in the former case, is a known feature \cite{Fatkhulla,VA,FRM}, explained by
the fact that the oscillations predicted by the VA are damped by the
radiation loss, which is not taken into regard by the VA based on simple
ansatz (\ref{ans}). A modification of the VA aimed to account for the
emission of radiation, in some approximation, was proposed \cite{KathSmyth},
but it is quite cumbersome even in the 1D setting.


\begin{figure}[tbp]
\begin{center}
\includegraphics[height=4.3785in, width=6.2396in]{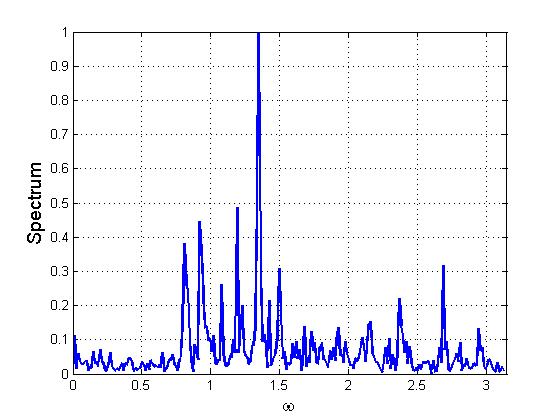}
\end{center}
\caption{(Color online) The power spectrum of small random perturbations
around the stable soliton trapped in the static lattice ($\protect%
\varepsilon =0$), for $V_{0}=0.25$ and $N=5.31$.}
\label{Pic_A_max_w}
\end{figure}

The VA predicts eigenfrequency $\omega _{0}$ of intrinsic oscillations of \
the weakly perturbed 2D soliton trapped in the static OL, see Eq. (\ref%
{omega0}). Because this feature may be important to understand the response
of the soliton to the periodic time modulation of the lattice, in terms of
possible resonances (see below), it is necessary to check whether the
presence of the eigenfrequency is confirmed by simulations of full equation (%
\ref{GPE}) with the static lattice, i.e., $\varepsilon =0$. To this end, in
Fig. \ref{Pic_A_max_w} we display the power spectrum of small oscillations
around the soliton caused by the addition of a small random perturbation to
it, for $V_{0}=0.25$ and $N=5.31$, which corresponds to $1-N/\tilde{N}_{\max
}\approx \allowbreak 0.155$. As mentioned above, in this case Eqs. (\ref{W0}%
) and (\ref{omega0}) predict the eigenfrequency to be $\omega _{0}\approx
1.35$. The spectrum in Fig. \ref{Pic_A_max_w} clearly shows the main peak
quite close to this point. Peaks corresponding to higher-order resonances
(see above) can also be recognized in the figure. To produce the spectrum
shown in Fig.\ref{Pic_A_max_w}, we simulated the evolution of the soliton up
to a very long time, $t=1200$, eliminating a contribution from a relatively
short initial stage, which featured a transient behavior.


\subsection{Stability diagrams}

\begin{figure}[tbp]
\begin{center}
\includegraphics[height=4.3785in, width=6.2396in]{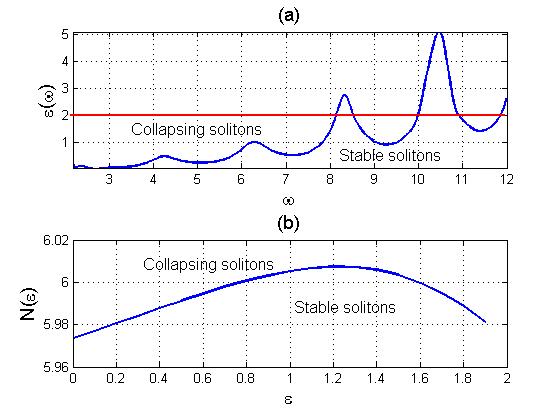}
\end{center}
\caption{(Color online) (a) The stability region in the plane of the
modulation parameters, $\protect\omega $ and $\protect\varepsilon $, for $%
V_{0}=0.25$ and $N=5.905$. The area relevant to the model with the
periodically modulated optical lattice corresponds to $\protect\varepsilon %
\leq 2$, where the time-dependent amplitude in Eq. (\protect\ref{GPE}), $%
1+\left( \protect\varepsilon /2\right) \cos (\protect\omega t)$, does not
change its sign. (b) The collapse threshold versus the modulation amplitude,$%
\ \protect\varepsilon $, for $V_{0}=0.25$ and $\protect\omega =4$.}
\label{Pic_epsilon(w)2}
\end{figure}

Results produced by the systematic numerical analysis of the stability of
the 2D soliton under the action of the ``lattice management" are collected
in Fig. \ref{Pic_epsilon(w)2}(a), which displays the stability region in the
plane of the modulation parameters, $\omega $ and $\varepsilon $, for fixed $%
V_{0}=0.25$ and $N=5.905$. Note that this norm again (like in the cases of
Figs. \ref{Pic_OneSol_PL} and \ref{Pic_sinhron}) slightly exceeds the
collapse threshold in the static model, which is given by Eq. (\ref{Nmax}).
The stability region in the plane of $\varepsilon $ and $N$, for the same OL
strength, $V_{0}=0.25$, as in Fig. \ref{Pic_epsilon(w)2}, and $\omega =4$,
is displayed in Fig. \ref{Pic_epsilon(w)2}(b). The latter plot explicitly
demonstrates the \emph{growth} of the collapse threshold, $N_{\max }$, with
the increase of the modulation amplitude. Note that not the entire parameter
region below the stability border in Fig. \ref{Pic_epsilon(w)2}(b)
corresponds to stable solitons; if $N$ is too small, the solitons are
unstable against decay, see below.


In Fig. \ref{Pic_epsilon(w)2}(a), the horizontal line denotes the maximum
possible value of the modulation amplitude in the model of the
time-modulated OL, $\varepsilon _{\max }=2$. A noteworthy fact is that the
stability region can extend up to this limit, i.e., $100\%$ modulation depth
(recall that, in the case when the stabilization of 2D solitons was provided
by the quasi-1D lattice, the stability limit corresponded to shallow
modulation \cite{Thawatchai2}). The results for $\varepsilon >2$ are
included too in Fig. \ref{Pic_epsilon(w)2}(a), as they may find a different
physical realization, corresponding to a superposition of moving OLs \cite%
{Staliunas:2007a} (as mentioned above).

A feature obvious in Fig. \ref{Pic_epsilon(w)2}(a) is a resonant-like
dependence of the stability border on $\omega $. This may be a manifestation
of the fundamental resonance at $\omega $ close to $\omega _{0}$ and
higher-order resonances at multiple values of $\omega $. As argued above,
the resonance may help to arrest the collapse, by forcing the vibrating
soliton to spend less time in the state where it is ``dangerously" narrow.
Since the value of the norm corresponding to Fig. \ref{Pic_epsilon(w)2}(a)
is close to $N_{\max }$, one may expect that the corresponding
fundamental-resonance frequency should be close to one given by Eq. (\ref%
{simple}), i.e., $\omega _{0}=2$, for $V_{0}=0.25$ [the same is given by
Eqs. (\ref{W0}) and (\ref{omega0}) in the limit of $N/\tilde{N}_{\max
}\rightarrow 0$]. Indeed, the picture in Fig. \ref{Pic_epsilon(w)2}(a) is
consistent with the expectation of the fundamental and higher-order
resonances at $\omega =n\omega _{0}$, $n=1,2,3,4$ (the picture also suggests
the existence of a very strong resonance close to $\omega =5\omega _{0}$,
but that one falls deeply into the unphysical region, $\varepsilon >2$).

\begin{figure}[tbp]
\begin{center}
\includegraphics[height=4.3785in, width=6.2396in]{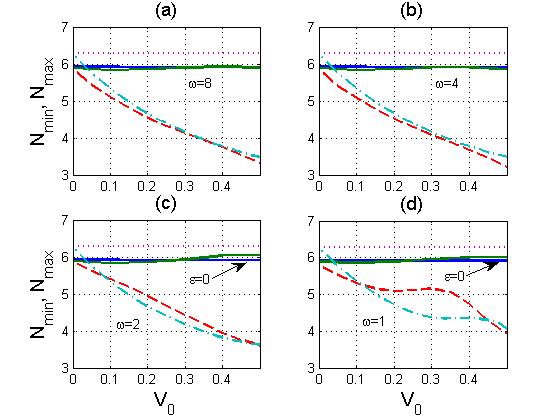}
\end{center}
\caption{(Color online) The dependence of the stability boundaries, $N_{\max
}$ and $\ N_{\min }$, on the OL strength, $V_{0}$, at $\protect\varepsilon %
=0.5$ and different values of the modulation frequency, $\protect\omega $.
Solid and dashed lines show, severally, $N_{\max }$ and $N_{\min }$ as found
from the simulations, while the dotted and dashed-dotted lines represent,
respectively, $N_{\max }$ and $N_{\min }$ as predicted by the variational
approximation. The arrows in (c) and (d) indicate values $N_{\max }$\ \ for
the stationary case, $\protect\varepsilon =0$.}
\label{Pic_n_MaxMin_V0}
\end{figure}

As mentioned above, the soliton whose norm is too small cannot stabilize
itself and decays into quasi-linear waves. Therefore, along with the upper
stability boundary, $N=N_{\max }$, that guarantees the absence of the
collapse, it is necessary to identify the lower one, $N=N_{\min }$, which
secures the stability of the soliton against the decay.\ Both boundaries, as
predicted by the VA and found from the direct simulations, are displayed in
Fig. \ref{Pic_n_MaxMin_V0}, in the form of dependences $N_{\max }(V_{0})$ and%
$\ N_{\min }(V_{0})$, at fixed $\varepsilon =0.5$ for several different
values of $\omega $.\ In each panel, the stability region is $N_{\min
}<N<N_{\max }$. 


\begin{figure}[tbp]
\begin{center}
\includegraphics[height=4.3785in, width=6.2396in]{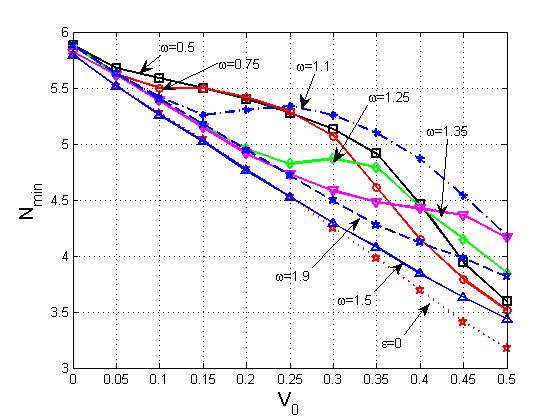}
\end{center}
\caption{(Color online) Dependencies $N_{\min }(V_{0})$ for $\protect%
\varepsilon =0.5$ and different modulation frequencies $\protect\omega $ in
the resonant area, $0.5<\protect\omega <2$ (see details in text). The curve
pertaining to $\protect\varepsilon =0$ (the static model) is included for
comparison.}
\label{Pic_nMin_V0_Details}
\end{figure}

We observe in Fig. \ref{Pic_n_MaxMin_V0} that dependences $N_{\min }(V_{0})$%
\ produced by the VA are in reasonable agreement with the results of direct
simulations of the GPE for modulation frequencies $\omega \geq 2$. However,
there is a conspicuous discrepancy between the VA and GPE for $\omega $
close to $1$. On the other hand, this range features strong resonances in
the perturbation spectrum, see Fig. \ref{Pic_A_max_w}. Therefore, it is
interesting to explore the $N_{\min }(V_{0})$\ dependence in region $%
0.5<\omega <2$, where the resonances may lead to decay of the soliton. These
results are displayed in Fig. \ref{Pic_nMin_V0_Details}, which shows that
plots $N_{\min }(V_{0})$\ essentially differ, in this region, from their
counterparts both in the static model (see the curve for $\varepsilon =0$\
in Fig. \ref{Pic_nMin_V0_Details}) and in the high-frequency region, $\omega
\geq 2$: actually, $N_{\min }$\ increases due to the resonant decay of the
solitons. The fact that the resonances are narrow enough, as seen in Fig. %
\ref{Pic_A_max_w}, may explain a non-monotonous character of dependences $%
N_{\min }(V_{0})$\ for $\omega =1.10$ and $1.25$.


\begin{figure}[tbp]
\begin{center}
\includegraphics[height=4.3785in, width=6.2396in]{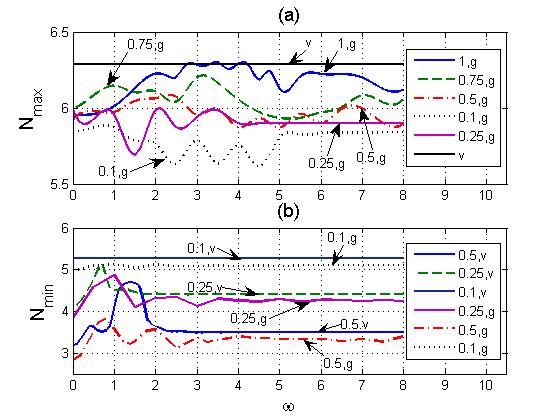}
\end{center}
\caption{(Color online) The dependence of the upper (a) and lower (b)
stability boundaries, $N_{\max }$ and $N_{\min }$, on modulation frequency $%
\protect\omega $ for $\protect\varepsilon =0.5$ and different values of OL
strength $V_{0}$, which are specified in the box. Labels ``v" and ``g"
pertain, severally, to the curves predicted by the variational approximation
and those found from direct simulations of Gross-Pitaevskii equation (%
\protect\ref{GPE}) (the VA predicts flat value $\tilde{N}_{\max }=2\protect%
\pi $, see text). The arrows identify those curves which may seem
indisnguishable in the black-and-white rendition of the figure.}
\label{pic_n_Min_pde_av}
\end{figure}

Further, dependences of $N_{\max }$ and $N_{\min }\ \allowbreak $on the
modulation frequency are displayed in Fig. \ref{pic_n_Min_pde_av}, for the
same modulation amplitude as above, $\varepsilon =0.5$, and several
different values of $V_{0}$. It is quite natural that dependences $N_{\min
}(\omega )$, as predicted by the VA and generated by direct simulations of
Eq. (\ref{GPE}), are close to each other for smaller values of the OL
strength, $V_{0}=0.1$ and $0.25$, while at $V_{0}=0.5$ the discrepancy
between them is considerable. Note non-monotonous behavior of $N_{\min
}(\omega )$\ in resonant zone $0.5<\omega <2$, which correlates to
peculiarities of dependences $N_{\min }(V_{0})$ in the same zone, cf. Fig. %
\ref{Pic_nMin_V0_Details}


\section{Conclusions}

We have studied the dynamics of 2D solitons in the model of BEC trapped in
the square-shaped OL (optical lattice) whose strength it subject to the
periodic time modulation. Being quite feasible for the experimental
implementation, the model belongs to a broad class of schemes of the
periodic management of solitons \cite{book}. By means of the VA (variational
approximation) and direct systematic simulations of the underlying
Gross-Pitaevskii equation, we have identified stability regions for the
solitons in the parameter space of the model, including both maximum and
minimum values of the norm, as summarized in Figs. \ref{Pic_epsilon(w)2}, %
\ref{Pic_n_MaxMin_V0}, \ref{Pic_nMin_V0_Details} and \ref{pic_n_Min_pde_av}.
A remarkable feature demonstrated by these results is that the stability
limit may reach the maximum ($100\%$) modulation depth. It is noteworthy too
that an \emph{increase} of the collapse threshold in predicted in comparison
with its classical value in the static situation, which corresponds to the
norm of the Townes soliton. The stability borders predicted by the VA are
found to be in reasonable agreement with the numerical results. In the plane
of the modulation frequency and amplitude, the stability boundary features a
salient resonant structure, which may also be qualitatively explained by
means of the VA. In particular, the minimum norm necessary for the stability
of the soliton demonstrates strong variations in the region of low
modulation frequencies, due to the possibility of the decay of the soliton
enhanced by the resonance.

The analysis reported in this work can be extended in several directions. It
may be interesting to study interactions between the solitons in this
setting, and identify stability limits for vortex solitons, as well as for
2D gap solitons in the model combining the repulsive nonlinearity and
lattice management. Another extension may be made in the direction of an
anisotropic lattice management, i.e., applying time modulations shifted by $%
\pi$ to the two 1D sublattices.

\section*{Acknowledgements}

The work of G.B. is partially supported by CONACyT grant No. 47220. The work
of B.A.M. was supported, in a part, by the Israel Science Foundation through
the Center-of-Excellence grant No. 8006/03.


\begin{thebibliography}{99}
\bibitem{Review} B. A. Malomed, D. Mihalache, F. Wise, and L. Torner, J.
Optics B: Quant. Semics. Opt. \textbf{7}, R53 (2005).

\bibitem{experiment} K. E. Strecker, G. B. Partridge, A. G. Truscott, and F.
G. Hulet, Nature \textbf{417}, 150 (2002); L. Khaykovich, F. Schreck, G.
Ferrari, T. Bourdel, J. Cubizolles, L. D. Carr, Y. Castin, and C. Salomon,
Science \textbf{296}, 1290 (2002); S. L. Cornish, S. T. Thompson, and C. E.
Wieman, Phys. Rev. Lett. \textbf{96}, 170401 (2006).

\bibitem{BBB} B. B. Baizakov, B. A. Malomed, and M. Salerno, Europhys. Lett.
\textbf{63}, 642 (2003); J. Yang and Z. H. Musslimani, Opt. Lett. \textbf{28}%
, 2094 (2003).

\bibitem{BBB2} B. B. Baizakov, B. A. Malomed and M. Salerno, Phys. Rev. A
\textbf{70}, 053613 (2004).

\bibitem{BBB3} B. B. Baizakov, B. A. Malomed and M. Salerno, in: \textit{%
Nonlinear Waves: Classical and Quantum Aspects}, ed. by F. Kh. Abdullaev and
V. V. Konotop, pp. 61-80 (Kluwer Academic Publishers: Dordrecht, 2004); also
available at http://rsphysse.anu.edu.au/\symbol{126}asd124/Baizakov\_2004%
\_61\_NonlinearWaves.pdf.

\bibitem{Kasprzak:2006a} J. Kasprzak, M. Richard, S. Kundermann, A. Baas, P.
Jeambrun, J. M. J. Keeling, F. M. Marchetti, M. H. Szymanska, R. Andre, J.
L. Staehli, V. Savona, P. B. Littlewood, B. Deveaud, and L. S. Dang, Nature
\textbf{443}, 409 (2006).

\bibitem{Barcelona} D. Mihalache, D. Mazilu, F. Lederer, Y. V. Kartashov,
L.-C. Crasovan, and L. Torner, Phys. Rev. E \textbf{70}, 055603(R) (2004).

\bibitem{gap1D} O. Zobay, S. P\"{o}tting, P. Meystre, and E. M. Wright,
Phys. Rev. A \textbf{59}, 643 (1999); F. Kh. Abdullaev, B. B. Baizakov, S.
A. Darmanyan, V. V. Konotop, and M. Salerno, \textit{ibid}. \textbf{64},
043606 (2001); A. Trombettoni, and A. Smerzi, Phys. Rev. Lett. \textbf{86},
2353 (2001); G. L. Alfimov, V. V. Konotop, and M. Salerno, Europhys. Lett.
\textbf{58}, 7 (2002).

\bibitem{GSstability} K. M. Hilligs\o e, M. K. Oberthaler, and K.-P.
Marzlin, Phys. Rev. A \textbf{66}, 063605 (2002); D. E. Pelinovsky, A. A.
Sukhorukov, and Y. S. Kivshar, Phys. Rev. E \textbf{70}, 036618 (2004).

\bibitem{Eiermann:2004a} B. Eiermann, Th. Anker, M. Albiez, M. Taglieber, P.
Treutlein, K.-P. Marzlin, and M. K. Oberthaler, Phys. Rev. Lett. \textbf{92}%
, 230401 (2004).

\bibitem{OliverMorsch:2006a} O. Morsch, M. Oberthaler, Rev. Mod. Phys. 78,
179215 (2006).

\bibitem{gap2D} B. B. Baizakov, V. V. Konotop, and M. Salerno, J. Phys. B:
At. Mol. Opt. Phys. \textbf{35}, 5105 (2002); P. J. Y. Louis, E. A.
Ostrovskaya, C. M. Savage, and Y. S. Kivshar, Phys. Rev. A \textbf{67},
013602 (2003); E. A. Ostrovskaya and Y. S. Kivshar, Opt. Exp. \textbf{12},
19 (2004); Phys. Rev. Lett. \textbf{93}, 160405 (2004); H. Sakaguchi and B.
A. Malomed, J. Phys. B \textbf{37}, 2225 (2004); A. Gubeskys, B. A. Malomed,
I. M. Merhasin, Phys. Rev. A \textbf{73}, 023607 (2006).

\bibitem{semi-gap} B. B. Baizakov, B. A. Malomed, and M. Salerno, Eur. Phys.
J. D \textbf{38}, 367 (2006).

\bibitem{HidetsuguSakaguchi:2006a} H. Sakaguchi, B. A. Malomed, Phys. Rev. E
74, 026601 (2006).

\bibitem{DDiso} P. Pedri and L. Santos, Phys. Rev. Lett. \textbf{95}, 200404
(2005).

\bibitem{DDaniso} I. Tikhonenkov, B. A. Malomed, and A. Vardi, Phys. Rev.
Lett. \textbf{100}, 090406 (2008).

\bibitem{FRM} H. Saito and M. Ueda, Phys. Rev. Lett. \textbf{90}, 040403
(2003); F. Kh. Abdullaev, J. G. Caputo, R. A. Kraenkel, and B. A. Malomed,
Phys. Rev. A \textbf{67}, 013605 (2003); G. D. Montesinos, V. M. P\'{e}%
rez-Garc\'{\i}a, and H. Michinel, Phys. Rev. Lett. \textbf{92}, 133901
(2004).

\bibitem{Isaac} I. Towers and B. A. Malomed, J. Opt. Soc. Am. \textbf{19},
537 (2002).

\bibitem{Staliunas:2004a} K. Staliunas, S. Longhi, and G. J. de Valc\'{a}%
rcel, Phys. Rev. A 70, 011601(R) (2004).

\bibitem{Warsaw} M. Trippenbach, M. Matuszewski, and B. A. Malomed,
Europhys. Lett. \textbf{70}, 8 (2005); M. Matuszewski, E. Infeld, B. A.
Malomed, and M. Trippenbach, Phys. Rev. Lett. \textbf{95}, 050403 (2005).

\bibitem{Mason} V. A. Brazhnyi and V. V. Konotop, Phys. Rev. A \textbf{72},
033615 (2005); M. A. Porter, M. Chugunova and D. E. Pelinovsky, Phys. Rev. E
74, 036610 (2006).

\bibitem{FRM-in-trap} P. G. Kevrekidis, G. Theocharis, D. J. Frantzeskakis,
and B.A. Malomed, Phys. Rev. Lett. \textbf{90}, 230401 (2003).D. E.
Pelinovsky, P. G. Kevrekidis, and D. J. Frantzeskakis, \textit{ibid}.
\textbf{91}, 240201 (2003); F. Kh. Abdullaev, R. M. Galimzyanov, M. Brtka,
and R. A Kraenkel, J. Phys. B: At. Mol. Opt. Phys. \textbf{37} (2004) 3535
(2004).

\bibitem{book} B. A. Malomed. \textit{Soliton Management in Periodic Systems}
(Springer:\ New York, 2006).

\bibitem{Itin} A. Itin, T. Morishita, and S. Watanabe, Phys. Rev. A 74,
033613 (2006).

\bibitem{Dalfovo} C. Tozzo, M. Kr\"{a}mer, and F. Dalfovo, Phys. Rev. A 72,
023613 (2005); M. Kr\"{a}mer, C. Tozzo, and F. Dalfovo, Phys. Rev. A \textbf{%
71}, 061602(R) (2005).

\bibitem{Thawatchai} T. Mayteevarunyoo and B. A. Malomed, Phys. Rev. A
\textbf{74}, 033616 (2006).

\bibitem{experiment-OL} C. Schori, T. St\"{o}ferle, H. Moritz, M. K\"{o}hl,
and T. Esslinger, Phys. Rev. Lett. \textbf{93}, 240402 (2004).

\bibitem{trap-modulation} J. J. G. Ripoll and V. M. P\'erez-Garc\'ia, Phys.
Rev. A 59, 2220 (1999); J. J. Garc\'ia-Ripoll, V. M. P\'erez-Garc\'ia, and
Pedro Torres, Phys. Rev. Lett. \textbf{83}, 1715 (1999); B. Baizakov, G.
Filatrella, B. Malomed, and M. Salerno, Phys. Rev. E \textbf{71}, 036619
(2005); P. Engels, C. Atherton, and M. A. Hoefer, Phys. Rev. Lett. \textbf{98%
}, 095301 (2007); Yu. Kagan and L. A. Manakova, Phys. Rev. A \textbf{76},
023601 (2007).

\bibitem{Faraday-theory} K. Staliunas, S. Longhi, and G. J. de Valc\'{a}%
rcel, Phys. Rev. Lett. \textbf{89}, 210406 (2002); Phys. Rev. A \textbf{70},
011601(R) (2004).

\bibitem{Engels} P. Engels, C. Atherton, and M. A. Hoefer, Phys. Rev. Lett.
\textbf{98}, 095301 (2007).

\bibitem{Randy} H. Saito, R. G. Hulet, and M. Ueda, Phys. Rev. A 76, 053619
(2007); H. Susanto, G. K. Kevrekidis, B. A. Malomed, and F. Kh. Abdullaev,
Phys. Lett. \textbf{372}, 1631 (2008).

\bibitem{Thawatchai2} T. Mayteevarunyoo, B. A. Malomed, and M. Krairiksh,
Phys. Rev. A \textbf{76}, 053612 (2007).

\bibitem{Staliunas:2007a} K. Staliunas, R. Herrero, and G. J. de Valc\'{a}%
rcel, Phys. Rev. A \textbf{75}, 011604(R) (2007).

\bibitem{Staliunas:2006a} K. Staliunas, R. Herrero, and G. J. de Valc\'{a}%
rcel, Phys. Rev. E \textbf{73}, 065603(R) (2006).

\bibitem{Gena} G. Burlak and A. Klimov, Phys. Lett. A \textbf{369}, 510
(2007).

\bibitem{Salerno} B. B. Baizakov and M. Salerno, Phys. Rev. A \textbf{69},
013602 (2004).

\bibitem{Desaix} M. Desaix, D. Anderson, M. Lisak, J. Opt. Soc. Am. B
\textbf{8}, 2082 (1991).

\bibitem{VA} V. M. P\'{e}rez-Garc\'{\i}a, H. Michinel, J. I. Cirac, M.
Lewenstein, and P. Zoller, Phys. Rev. A \textbf{56}, 1424 (1997); B. A.
Malomed, in Progress in Optics,\textit{\ }vol. 43, p. 71, ed. by E. Wolf
(Amsterdam: North-Holland, 2002).

\bibitem{Berge} L. Berg\'{e}, Phys. Rep. \textbf{303}, 259 (1998).

\bibitem{Landau} L. D. Landau and E. M. Lifshitz, \textit{Mechanics}, 3rd
ed. (Pergamon Press: Oxford, England, 1976).

\bibitem{Fatkhulla} F. Kh. Abdullaev and J. G. Caputo, Phys. Rev. E \textbf{%
58}, 6637 (1998).

\bibitem{Press} W. H. Press, S. A. Teukovsky, W. T. Vetterling, and B.
P.Flannery, \textit{Numerical recipes in C++} (Cambridge University Press:
Cambridge, 2002).

\bibitem{KathSmyth} W. L. Kath and N. F. Smyth, Phys. Rev. E \textbf{51},
1484 (1995).
\end{thebibliography}
\end{document}